\documentclass[showpacs,twocolumn,amsmath,amssymb]{revtex4}
\usepackage{graphicx}
\usepackage{psfrag} 
\usepackage{epsfig}    
\usepackage{rotating}  

\begin{document}

\title{Dynamics of Vacillating Voters}
\author{R. Lambiotte$^{1,2}$}
\author{S. Redner$^{3}$}
\affiliation{$^{1}$GRAPES, Universit\'e de Li\`ege, Sart-Tilman, B-4000
  Li\`ege, Belgium \\
  $^2$ INMA, Universit\'e catholique de Louvain,  4 avenue Georges Lemaitre,
B-1348 Louvain-la-Neuve, Belgium \\
$^{3}$Center for Polymer Studies and Physics  Department, Boston University, Boston,
MA~ 02215 USA}


\begin{abstract}
  We introduce the vacillating voter model in which each voter consults two
  neighbors to decide its state, and changes opinion if it disagrees with
  either neighbor.  This irresolution leads to a global bias toward zero
  magnetization.  In spatial dimension $d>1$, anti-coarsening arises in which
  the linear dimension $L$ of minority domains grows as $t^{1/(d+1)}$.  One
  consequence is that the time to reach consensus scales exponentially with
  the number of voters.

\end{abstract}

\maketitle

The voter model \cite{L99} gives an appealing, albeit idealized, description
for the opinion dynamics of a socially interacting population.  In this
model, each node of a graph is occupied by a voter that has one of two
opinions, $\uparrow$ or $\downarrow$.  The population evolves by: (i) picking
a random voter; (ii) the selected voter adopts the state of a randomly-chosen
neighbor; (iii) repeating these steps {\it ad infinitum} or until a finite
system necessarily reaches consensus.  Descriptively, each voter has no self
confidence and follows one of its neighbors.  With this dynamics, a voter
chooses a state with a probability equal to the fraction of neighbors in that
state, a feature that renders the voter model soluble in all dimensions
\cite{L99,K02}.

In this work, we investigate a variation that we term the {\em vacillating\/}
voter model.  By vacillating, we mean that a voter very much lacks confidence
in its state.  In an update, if a voter happens to select a random neighbor
of the same persuasion, the voter is still not convinced that this state is
right.  Thus the voter selects another random neighbor and adopts this state.
This vacillation causes a voter to change state with a larger probability
than the fraction of disagreeing neighbors, and leads to a bias toward the
zero-magnetization state in which there are equal densities of voters of each
type.

\begin{figure}[ht]
\includegraphics[width=0.4\textwidth]{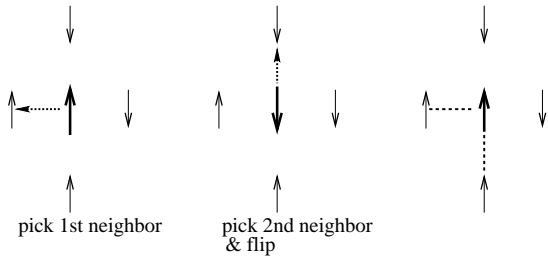}
\caption{Illustration of an update for the vacillating voter on the square
  lattice (left and middle).  For the configuration on the right, the central
  voter flips with probability 5/6 because out of the 6 ways of selecting two
  neighbors, only one choice leads to both neighbors agreeable (dashed).}
\label{model}      
\end{figure}

Thus vacillation inhibits consensus, but due to a different mechanism than
that in the prototypical Axelrod model \cite{A}, the bounded compromise model
\cite{W} and its variants \cite{VR}.  For these latter models, consensus is
hindered because of the absence of interaction whenever two agents become
sufficiently incompatible.  For vacillating voters, it is individual
uncertainty that forestalls consensus.  The vacillating voter model also
differs from models that incorporate ``contrarians'' \cite{galam} because
voters still try to imitate their neighbors.

The update steps in the vacillating voter model are:
\begin{enumerate}
\itemsep -1pt
\item Pick a random voter.

\item The voter picks a random neighbor.  If the neighbor disagrees with the
  voter, the voter changes state.

\item If the neighbor and the voter agree, the voter picks another random
  neighbor and adopts its state.
  
\item Repeat steps 1 and 2 {\em ad infinitum} or until consensus is reached.
\end{enumerate}

For example, the probability that a vacillating voter on the square lattice
flips is $0,\frac{1}{2},\frac{5}{6}$, and 1, respectively, when the number of
anti-aligned neighbors is 0, 1, 2, and $\geq 3$ (Fig.~\ref{model}).  In
contrast, for the classic voter model, the flip probability is $\frac{k}{4}$,
where $k$ is the number of neighbors of the opposite opinion.  We now explore
the consequences of this vacillation on voter dynamics.

Consider first the mean-field limit.  Here the density $x$ of $\uparrow$
voters obeys the rate equation
\begin{eqnarray}
\dot x &=& -x \left[1-x^2\right] + (1-x)\left[1-(1-x)^2\right]\nonumber\\
&=& x(1-x)(1-2x).
\end{eqnarray}
The first term on the right accounts for the loss of $\uparrow$ voters in
which a $\uparrow$ voter is first picked (factor $x$), and then the
neighborhood cannot consist of two $\uparrow$ voters (factor $1-x^2$).
Similarly, in the second (gain) term, a $\downarrow$ voter is first picked,
and then the neighborhood must contain at least one $\uparrow$ voter.  The
factorized form shows that there are unstable fixed points at $x=0,1$ and a
stable fixed point at $x=1/2$.  Thus a population is driven to the
zero-magnetization state.

However, because consensus is the only absorbing state of the stochastic
dynamics, a finite population ultimately reaches consensus.  To characterize
the evolution to this state, we first study the exit probability
$\mathcal{E}_n$, defined as the probability that a population of $N$ voters
ultimately reaches $\uparrow$ consensus when there are initially $n$
$\uparrow$ voters.  Then $\mathcal{E}_n$ obeys the backward equation
\cite{fpp}
\begin{equation}
\label{E}
\mathcal{E}_n=w_{n\to n+1} \,\mathcal{E}_{n+1} + w_{n\to n-1} \,\mathcal{E}_{n-1}
+ w_{n\to n} \,\mathcal{E}_{n},
\end{equation}
where $w_{n\to m}$ is the probability for the transition from the state with
$n$ $\uparrow$ voters to $m$ $\uparrow$ voters in an update.  This equation
expresses the probability to exit from $n$ as the probability to take one
step (the factors $w$) times the probability to exit from the point reached
after one step.  In the large-$n$ limit, we write $x=n/N$, and the transition
probabilities become
\begin{eqnarray*}
w_{n\to n+1}&=& (1-x)\left[1-(1-x)^2\right]\\
w_{n\to n-1}&=& x(1-x^2)\\
w_{n\to n}&= &x^3+ (1-x)^3.\\
\end{eqnarray*}

\begin{figure}[ht]
\includegraphics[width=0.45\textwidth]{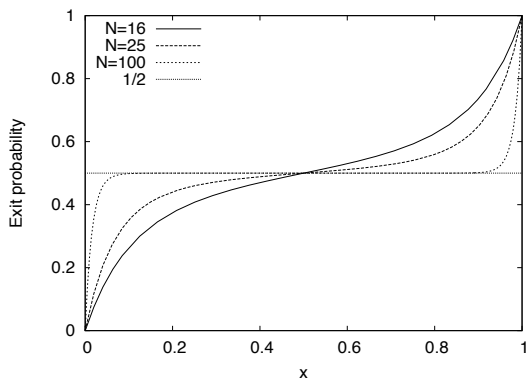}
\vskip -.3in
\caption{Exit probability $\mathcal{E}(x)$ versus the density of
$\uparrow$ voters $x$ for the case $N=16$, $N=25$ and $N=100$.}
\label{exit}      
\end{figure}

Substituting these in \eqref{E}, writing $\mathcal{E}_{n\pm
  1}\to\mathcal{E}(x\pm \delta x)$, and expanding to second order in $\delta
x$, gives
\begin{equation}
  \frac{3 x(1-x)}{2N} \frac{\partial^2\mathcal{E}}{\partial x^2} 
  + x(1-x)(1-2x) \frac{\partial\mathcal{E}}{\partial x}=0,
\end{equation}
with solution
\begin{equation}
\label{exitMF}
\mathcal{E}(x) = \int_{-1/2}^{x-1/2} e^{2Ny^2/3}\, dy\Bigg/ 
\int_{-1/2}^{1/2} e^{2Ny^2/3}\, dy.
\end{equation}
Notice that $\mathcal{E}(x)$ approaches the constant value $1/2$ for
increasing $N$ (Fig.~\ref{exit}), reflecting the bias towards the
zero-magnetization state.  Almost all initial states are driven to the
potential well at $x=1/2$, so that the exit probability becomes independent
of the initial density of $\uparrow$ voters.

Similarly, we study the time to reach consensus as a function of the initial
composition of voters.  Let $t_n$ denote the time to reach consensus (either
all $\uparrow$ or all $\downarrow$) when starting with $n$ $\uparrow$ voters
in a population of $N$ voters.  Similar to \eqref{E}, $t_n$ obeys the
backward equation \cite{fpp}
\begin{equation}
\label{T}
t_n= \delta t+ w_{n\to n+1} \,t_{n+1}  +w_{n\to n-1} \,t_{n-1}+ w_{n\to n} \,t_{n},
\end{equation}
where $\delta t=1/N$ is the time elapsed in an update.  In the large-$n$
limit, this equation becomes
\begin{equation}
\frac{3 x(1-x)}{2N}  \frac{\partial^2 t}{\partial x^2} 
+ x(1-x)(1-2x)  \frac{\partial t}{\partial x}=-1.
\end{equation}
The formal solution is again elementary, but the result can no longer be
expressed in closed form.  The main result is that the consensus time scales
as $e^{a N}$, with $a$ a constant of order 1.  In contrast to the classical
voter model, the global bias drives the system into a potential well that
must be surmounted to reach consensus.  Thus the consensus time is
anomalously long.

In one dimension, a voter changes its opinion if at least one of its
neighbors is in disagreement.  For example, a $\uparrow$ voter flips with
rate 1 if the neighborhood configurations are $\uparrow \uparrow \downarrow$,
$\downarrow \uparrow \uparrow$, and $\downarrow \uparrow \downarrow$.  As an
amusing side-note, this dynamics is equivalent to rule 178 of the
one-dimensional cellular automaton \cite{wolfram}, except that this rule is
implemented asynchronously in the vacillating voter model.  In the framework
of the Ising-Glauber model \cite{IG}, the flip rate of a voter at site $i$,
whose states are now represented by $\sigma_i=\pm 1$, is
\begin{eqnarray}
\label{w}
w(\{\sigma\}\!\!\rightarrow\!\!\{\sigma'\}_i) \!=\!
 -\frac{\left[\sigma_{i}(\sigma_{i+1}\!+\!
 \sigma_{i-1}) \!+\! \sigma_{i-1} \sigma_{i+1} \! -\! 3\right]}{4}, 
\end{eqnarray}
with $\{\sigma\}$ denoting the state of all voters and $\{\sigma'\}_i$ the
state where the $i^{\rm th}$ voter flips.  The first two terms correspond to
conventional Glauber kinetics, but as mentioned parenthetically in
Ref.~\cite{IG}, the presence of the $\sigma_{i-1}\sigma_{i+1}$ term couples
the rate equation for the mean spin to 3-body terms and the model is not
exactly soluble.

The mean spin, $s_j\equiv\langle\sigma_j\rangle=\sum_{\{\sigma\}} \sigma_j
P(\{\sigma\};t)$ evolves according to
\begin{eqnarray}
\label{st}
\frac{\partial s_j}{\partial t}&=& \sum_{\{\sigma\}}\sigma_j\Big[\sum_i w(\{\sigma'\}_i
\rightarrow \{\sigma\})\, P(\{\sigma'\}_i;t) \nonumber \\
&~&~~~~~~~~-w(\{\sigma\} \rightarrow \{\sigma'\}_i) \, P(\{\sigma\};t)\Big] ,
\end{eqnarray}
which reduces to, after straightforward but tedious steps,
\begin{eqnarray}
\label{eqS}
\frac{\partial s_j}{\partial t}  =  \frac{1}{2}\left(s_{j+1} +  s_{j-1} +  
\langle\sigma_{j-1}\sigma_j\sigma_{j+1}\rangle  - 3s_j\right).
\end{eqnarray}

In a similar spirit, the rate equation for the nearest-neighbor correlation
function, $\langle\sigma_j \sigma_{j+1}\rangle$, is
\begin{eqnarray}
\label{ct}
\frac{\partial \langle\sigma_j \sigma_{j\!+\!1}\rangle}{\partial t} &=&
\frac{1}{2}\left[\langle \sigma_{j\!-\!1}(\sigma_j\!+\!\sigma_{j\!+\!1})\rangle+
\langle (\sigma_j\!+\!\sigma_{j\!+\!1})\sigma_{j\!+\!2}\rangle\right]~~~~~\nonumber \\
&~&~~~~~~~~ +1 -3\langle\sigma_j\sigma_{j+1}\rangle
\end{eqnarray}

We can simplify Eq.~\eqref{ct} by considering domain walls---nearest-neighbor
anti-aligned voters---whose density is given by $\rho=(1-\langle
\sigma_i\sigma_{i+1}\rangle)/2$.  According to the flip rate in
Eq.~\eqref{w}, an isolated domain wall diffuses freely, just as in the pure
voter model.  However, when two domain walls are adjacent, they annihilate
with probability 1/3 or one hops away from the other with probability 2/3.
This process is isomorphic to single-species annihilation, $A+A\to 0$, but
with a reduced reaction rate compared to freely diffusing reactants because
of the nearest-neighbor repulsion.  The domain wall density still
asymptotically decays as $t^{-1/2}$ with an amplitude that depends on the
magnitude of the repulsion.

Because domain walls are widely separated at long times, the second-neighbor
correlation function is
\begin{eqnarray*}
\langle\sigma_{j} \sigma_{j+2}\rangle&=& +\text{prob(0 or 2 walls
  between\ }j\ \text{and\ } j\!+\!2)\\
&~&- \text{prob(1 wall between\ }j\ \text{and\ } j\!+\!2)\\
&\approx& 1-2\rho.
\end{eqnarray*}
Using the approximation of widely separated domain walls, $\langle\sigma_{j}
\sigma_{j+2}\rangle \approx \langle\sigma_{j} \sigma_{j+1}\rangle\equiv m_2$,
and the rate equation for nearest-neighbor correlation function $m_2$ becomes
$\frac{\partial m_2}{\partial t} = 1 - m_2$, with solution
\begin{eqnarray}
m_2(t) =  1 + \left[m(0)^2-1\right]\, e^{-t} .
\end{eqnarray}
Here we chose the uncorrelated initial condition, so that $m_2(0)=m(0)^2$,
where $m(0)\equiv\langle s_j(0)\rangle $ is the average magnetization at
$t=0$.

\begin{figure}[ht]
\includegraphics[width=0.45\textwidth]{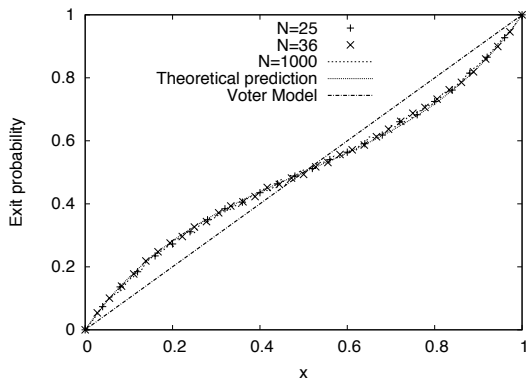}
\vskip -0.3in
\caption{Exit probability $\mathcal{E}(x)$ as a function of the initial
  density of $\uparrow$ voters $x$ for a one dimensional system composed of
  25, 36 and 1000 voters respectively The voter model result,
  $\mathcal{E}(x)=x$, that follows from magnetization conservation is shown
  for comparison.}
\label{fig3}      
\end{figure}

Let us now return to the rate equation \eqref{eqS} for the mean spin.  For a
spatially homogeneous system, $\langle s_j\rangle$ are all identical and the
magnetization is $m\equiv \langle s_j\rangle$.  Also, we follow Ref.~
\cite{MR} and decouple the 3-spin correlation function as $\langle
\sigma_{j-1} \sigma_j \sigma_{j+1}\rangle \approx m m_2$.  Then by averaging
over all sites, the rate equation equation \eqref{eqS} becomes
\begin{eqnarray}
\frac{\partial m}{\partial t}  &=&   \frac{1}{2}(m\, m_2 -m)
=  \frac{m}{2}   e^{-t} (m(0)^2-1)\,,
\end{eqnarray}
whose solution, for the initial condition $m(0)$, is
\begin{eqnarray}
  m(t)  &=&  m(0)\, e^{\frac{1}{2} (1-e^{-t}) (m(0)^2-1)}~.
\end{eqnarray}
Thus we obtain a non-trivial relation between final magnetization $m(\infty)$
and $m(0)$
 \begin{eqnarray}
 \label{mInf}
m(\infty)  =  m(0)\, e^{\frac{1}{2}  (m(0)^2-1)}~.
\end{eqnarray}
Since the density of $\uparrow$ voters is $x=(1+m)/2$, while
$m(\infty)=2\mathcal{E}(x)-1$, the exit probability $\mathcal{E}(x)$ becomes
 \begin{eqnarray}
 \label{exit1D}
\mathcal{E}(x)  =  \frac{1}{2} \left[(2x-1) e^{2x(x-1)} +1 \right].
\end{eqnarray}
This result is in excellent agreement with our simulation results
(Fig.~\ref{fig3}).  For small systems ($N=25$ and $36$), we directly measure
the probability $\mathcal{E}(n)$ that the population ultimately reaches a
$\uparrow$ consensus when there are initially $n$ $\uparrow$ voters and
averaged over 5000 realizations of the dynamics.  We also verified
Eq.~(\ref{exit1D}) for large systems ($N=1000$ nodes) by a different approach
that avoids the need to measure $\mathcal{E}(n)$ directly by simulating until
ultimate consensus.  Instead, we run the dynamics up to $1000$ time steps and
measure the magnetization at this time.  We then average over 200
realizations of the process to obtain $m(\infty)$ and finally obtain
$\mathcal{E}(x)$ from $\mathcal{E}(x)=(1+m(\infty))/2$.  We again find
excellent agreement with our prediction \eqref{exit1D}.  

\begin{figure}[ht]
  \includegraphics[width=0.35\textwidth]{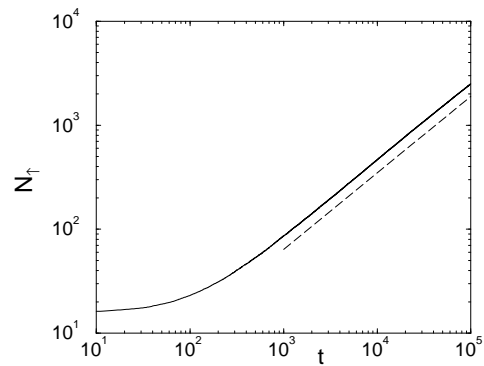}
  \caption{Double logarithmic plot of the number of $\uparrow$ voters versus
    time on the square lattice starting from a $4\times 4$ square of
    $\uparrow$ voters in a background of $\downarrow$ voters.  }
\label{a-vs-t}
\end{figure}

The vacillating voter model in greater than one dimension has the new
qualitative feature that small minority domains tend to grow.  This
anti-coarsening is a manifestation of the bias toward the zero-magnetization
state.  To appreciate how this anti-coarsening arises, consider a circular
two-dimensional island domain of $\uparrow$ voters of linear dimension $L$
and area $A$ in a sea of $\downarrow$ voters.  For large $L$, each voter at
the interface has the same local environment, so that there is no
environmental bias.  However, there are slightly more $\downarrow$ voters
just outside the circle that $\uparrow$ voters just inside.  In a time of the
order of $\delta t\sim L$ each interface voter is updated once, on average,
so that the island area increases by an amount $\delta A$ that is of the
order of the difference in the number of $\uparrow$ and $\downarrow$ voters
at the interface.  Thus $\frac{\delta A}{\delta t}\sim \frac{1}{L}$, which
gives $L\sim t^{1/3}$.  In $d$ dimensions, this same reasoning gives $L\sim
t^{1/(d+1)}$.  We probed for this anti-coarsening by simulating the evolution
of an initial small square domain of $\uparrow$ voters in a $\downarrow$
background in two dimensions (Fig.~\ref{a-vs-t}).  Although such domains do
not remain contiguous, the data suggest that the number, or occupied area, of
$\uparrow$ voters grows as $t^{\alpha}$, with $\alpha$ around 0.73, in
reasonable agreement with our expectation $\alpha=2/3$.

\begin{figure}[ht]
\includegraphics[width=0.45\textwidth]{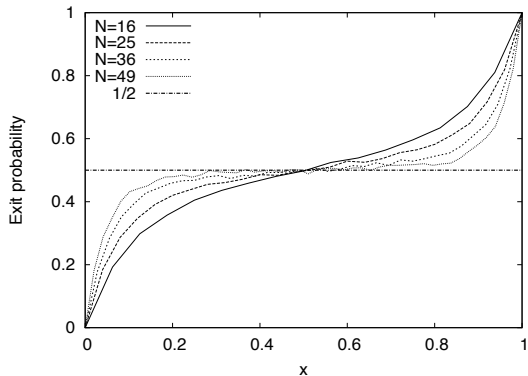}
\vskip -0.3in
\caption{Exit probability $\mathcal{E}(x)$ as a function of the initial
  density of $\uparrow$ voters $x$ for a square lattice of 16, 25,
  36 and 49 voters, respectively, with periodic boundary conditions.  }
\label{fig5}      
\end{figure}

\begin{figure}[ht]
  \includegraphics[width=0.5\textwidth]{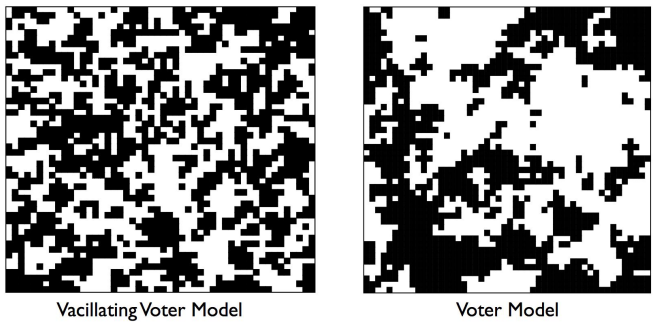}
  \caption{Snapshots of the vacillating (left) and pure (right) voter model
    on a $50 \times 50$ lattice starting with a random zero-magnetization
    state after $100$ time steps.  The correlation function $C_1$ equals
    $0.31$ (left) and $0.59$ (right) respectively.  }
\label{fig6}
\end{figure}

A system with non-zero initial magnetization is therefore again drawn to the
attractor where the density $x$ of $\uparrow$ voters equals $1/2$ before
final consensus is eventually reached.  It is only for $x$ initially very
close to 0 or 1 that the system achieves consensus without first being drawn
to this attractor.  Thus the exit probability $\mathcal{E}(x)$ should be
nearly independent of $x$ for almost all $x$, just as in the mean-field
limit.  Simulations of the vacillating voter model on the square lattice
(Fig.~\ref{fig5}) confirm that $\mathcal{E}(x)$ approaches 1/2 for a
progressively wider range of $x$ as $L$ increases.  Simulations also show
that the correlation function $C_1\equiv \langle \sigma_{i,j}
\sigma_{i,j+1}\rangle$ does not approach 1 in the long-time limit, as in one
dimension or in the pure voter model in two dimensions.  Rather, $C_1$
reaches the stationary value $0.31$, so that domains of opposite opinions
coexist (Fig.~\ref{fig6}), and only a rare macroscopic fluctuation allows
consensus to be reached.

In summary, when vacillation is incorporated into the voter model, consensus
is inhibited but not prevented.  In the mean-field limit, the vacillation
drives a population away from consensus and toward the zero-magnetization
state.  A finite system ultimately achieves consensus only via a macroscopic
fluctuation that allows the system to escape this bias-induced potential
well.  Because of the bias, the probability to reach $\uparrow$ consensus is
essentially independent of the initial composition of the population.  In one
dimension, the system coarsens, albeit more slowly than in the pure voter
model because of the repulsion of neighboring domain walls, and the
probability to reach the final state of $\uparrow$ consensus has a
non-trivial initial state dependence.  In two and higher dimensions, domains
slowly anti-coarsen to drive the system to the zero-magnetization state.  The
overall behavior is qualitatively similar to that of the mean-field
vacillating voter model, and very different from the pure voter model.

\acknowledgments We gratefully acknowledge the support of the European
Commission Project CREEN FP6-2003-NEST-Path-012864 (RL), the ARC ``Large
Graphs and Networks'' (RL) and NSF grant DMR0535503 (SR), and the hospitality
of the Ettore Majorana Center where this project was initiated.

\end{document}